\documentclass[11pt]{article}
\usepackage{amsmath,amssymb,color}

\usepackage{amsfonts}

\newcommand{\hoch}[1]{$\, ^{#1}$}

\newcommand{\be}{\begin{equation}}
\newcommand{\ee}{\end{equation}}
\newcommand{\bea}{\setlength\arraycolsep{2pt} \begin{eqnarray}}
\newcommand{\eea}{\end{eqnarray}}

\def\ft#1#2{{\textstyle{\frac{\scriptstyle #1}{\scriptstyle #2} } }}
\def\fft#1#2{{\frac{#1}{#2}}}

\def\0{{\sst{(0)}}}
\def\1{{\sst{(1)}}}
\def\2{{\sst{(2)}}}
\def\3{{\sst{(3)}}}
\def\4{{\sst{(4)}}}
\def\5{{\sst{(5)}}}
\def\6{{\sst{(6)}}}
\def\7{{\sst{(7)}}}
\def\8{{\sst{(8)}}}
\def\sst#1{{\scriptscriptstyle #1}}

\begin{document}

\begin{flushright}
\hfill{KIAS-P11032}
\end{flushright}

\vspace{25pt}
\begin{center}
{\large {\bf Killing Spinors for the Bosonic String}}

\vspace{10pt}

H. L\"u\hoch{1,2} and Zhao-Long Wang\hoch{3}

\vspace{10pt}

\hoch{1}{\it China Economics and Management Academy\\
Central University of Finance and Economics, Beijing 100081, China}

\vspace{10pt}

\hoch{2}{\it Institute for Advanced Study, Shenzhen
University\\ Nanhai Ave 3688, Shenzhen 518060, China}

\vspace{10pt}

\hoch{3} {\it School of Physics, Korea Institute for Advanced Study,
Seoul 130-722, Korea}

\vspace{30pt}

\underline{ABSTRACT}
\end{center}

We obtain the effective action for the bosonic string with arbitrary
Yang-Mills fields, up to the $\alpha'$ order, in general dimensions.
The form of the action is determined by the requirement that the action
admit well-defined Killing spinor equations, whose projected
integrability conditions give rise to the full set of equations of
motion.  The success of the construction suggests that the hidden ``pseudo-supersymmetry" associated with the Killing spinor equations
may be a property of the bosonic string itself.

\vspace{15pt}

\thispagestyle{empty}





\newpage

One of the most important advances in quantum gravity was the
invention of string theory.  Quantum consistency in general requires
that string theories be supersymmetric in the critical dimension
ten.\footnote{Exotic tachyon-free string theory without
supersymmetry is also possible with appropriate discrete torsion
introduced in the partition function \cite{pol}.} With the discovery
of the AdS/CFT correspondence \cite{mald}, the application of string
theory goes beyond quantum gravity, but also provides new
understandings and techniques in non-perturbative quantum field
theories including nuclear and condensed matter physics.  However,
these low-energy arenas are necessarily non-supersymmetric, and
hence the hitherto abandoned bosonic string theory may provide an
alternative and useful approach.

One advantage of supersymmetry is that it provides a powerful
organizing tool for the non-perturbative region. In particular the
characteristic properties of the BPS solutions of supergravities,
such as the mass-charge relation, are expected to survive the
higher-order quantum corrections. The defining property for the BPS
solutions is that they admit Killing spinors. Killing spinors,
however, were introduced in ordinary Riemannian geometry which
predates supersymmetry.  Killing spinors in supergravities are
generalizations of those in the Riemannian geometry. They all have
an important defining property that certain projected integrability
conditions for the Killing spinor equations give rise to the full
set of equations of motion of the bosonic fields. This property is a
necessary condition for constructing supergravities, and conversely,
all supergravities have consistent Killing spinor equations. It is
natural to ask whether there exist non-supersymmetric theories other
than pure gravities that also admit well-defined Killing spinors.

Recently, an intriguing connection between the consistency of the
Kluza-Klein sphere reduction and Killing spinors was observed in
\cite{lpw}. Inspired by this and also the fact that it is consistent
to perform $S^3$ and $S^{D-3}$ reductions on the effective action of
the bosonic string \cite{clpred}, Killing spinor equations were
proposed for this action \cite{lpw}. It was shown that the projected
integrability condition gives rise precisely to the full set of
equations of motion.
For non-supersymmetric theories, such examples are uncommon
\cite{lw}. The only known non-trivial examples are pure gravity,
certain scalar-gravity theories \cite{fake} and the bosonic string.
This implies that the bosonic string has a non-trivial generalized
geometric structure arising from the ``pseudo-supersymmetry'', a
terminology we introduce here to refer to the hidden symmetry
associated with the existence of well-defined Killing spinor
equations in non-supersymmetric theories.  From the AdS/CFT point of
view, where the bulk gravity is classical, the pseudo-supersymmetric
effective action of the bosonic string can be put on an equal
footing as in supergravity.

The pseudo-supersymmetry obtained in \cite{lpw} is for the
tree-level action of the bosonic string. If the pseudo-supersymmetry
is indeed a generic stringy property, we would expect that the
Killing spinor equations should remain well-defined when
higher-order corrections are included. In this paper, we extend the
discussion of \cite{lpw} by considering the $\alpha'$ correction to
the low-energy effective action.  We start by first introducing
arbitrary Yang-Mills fields, which originate in diverse ways in
string theory.  For an example, wrapping the string on singular
cycles of the internal manifold can lead to gauge symmetry
enhancement. One can also consider hybrid mixing between the left-
and right-moving sectors, with the mismatching dimensions
compactified on some suitable self-dual lattices, {\it \`a la}
heterotic string \cite{het}. The effective action with arbitrary
Yang-Mills fields in $D$ dimensions is given by
\begin{equation}
{\cal L}_D= \sqrt{-g} \Big( R - \ft12(\partial\phi)^2 - \ft{1}{12}
e^{a\phi} G_\3^2 - \ft14 e^{\ft12a\phi} {\rm tr}'
F_\2^2\Big)\,,\label{genlag}
\end{equation}
where $a^2=8/(D-2)$ and the various form fields are given by
\begin{eqnarray}
G_\3&=&dB_\2 - \ft12 \omega_\3\,,\quad \omega_\3 = {\rm tr}'
(F_\2\wedge A_\1 - \ft13 A_\1\wedge A_\1\wedge A_\1)\,,\cr 
F_\2 &=& dA_\1 + A_\1\wedge A_\1\,,\quad dG_\3 = -\ft12{\rm tr}'
(F_\2\wedge F_\2)\,.
\end{eqnarray}
The Yang-Mills 1-forms are defined by $A_\1=A^I T_I$, where the
generators $T_I$ are anti-hermitian, obeying the Lie algebra $[T_I,
T_J]=f^K{}_{IJ} T_K$, and normalized in the fundamental
representation as ${\rm tr} (T_I T_J) \equiv \beta \delta_{IJ}$.
Then the trace ${\rm tr}'$ is defined by ${\rm tr}' = \fft{1}{\beta}
{\rm tr}$.  Note that for simplicity, we have set the Yang-Mills
coupling to unity.

We propose that the defining equations for the Killing spinors for
(\ref{genlag}) are given by
\begin{eqnarray}
D_M\epsilon + \ft{1}{96} e^{\fft12a\phi}\Big(a^2 \Gamma_M
\Gamma^{NPQ} - 12 \delta_{M}^N\Gamma^{PQ}\Big) G_{NPQ}\,\eta
&=&0\,,\label{ks1}\\
\Gamma^M\partial_M \phi\, \eta + \ft{1}{12}a e^{\fft12a\phi}
\Gamma^{MNP} G_{MNP}\,\eta&=&0\,, \label{ks2}\\
\Gamma^{M_1M_2} F_{M_1M_2}\eta &=&0\,.\label{ks3}
\end{eqnarray}
When $D=10$, these are precisely the Killing spinor equations for
$D=10$, ${\cal N}=1$ supergravity \cite{Bvn} with Yang-Mills matter
multiplets. Let us now examine the consistency of these equations.
For Riemannian Killing spinors, satisfying $D_M\epsilon=0$, the
integrability condition is given by $[D_M, D_N]\epsilon=\ft14
R_{MNPQ}\Gamma^{PQ}\epsilon=0$. The projected integrability
condition, namely $\Gamma^M [D_M, D_N]\epsilon=\ft12
R_{MN}\Gamma^M\epsilon=0$, is satisfied by virtue of the Einstein
equation. The projected integrability conditions for the equations
(\ref{ks1})-(\ref{ks3}) are much more involved. We find that the
condition associated with $\Gamma^M [D_M, D_N]\eta$ is given by
\begin{eqnarray}
&&\Big[R_{MN}-\ft{1}{2} \partial_M\phi\partial_N\phi
-\ft14e^{a\phi}(G^2_{MN}-\ft{2}{3(D-2)} G^2 g_{MN})\cr
&&\qquad\qquad -\ft12e^{\fft12a\phi}{\rm tr}'(F^2_{MN} -
\ft{1}{2(D-2)}F^2 g_{MN}) \Big]\Gamma^{N} \eta \cr 
&&-\ft{ e^{\fft12a\phi} }{6(D-2)}(\nabla_N G_{M_1M_2M_3}+\ft34
F_{NM_1}F_{M_2M_3})\left(\Gamma_{M}\Gamma^{NM_1M_2M_3}
-2(D-2)\delta_M^{[N}\Gamma^{M_1M_2M_3]}\right)\eta \cr 
&&-\ft{e^{-\fft12a\phi}}{2(D-2)}\nabla_N \left( e^{a\phi}
G^N{}_{M_2M_3}\right)
\left(\Gamma_{M}\Gamma^{M_2M_3}-(D-2)\delta_M^{M_2}\Gamma^{M_3}\right)\eta
=0\,.\label{intcon1}
\end{eqnarray}
Acting with $\Gamma^N D_N$ on (\ref{ks2}) and (\ref{ks3}) gives rise to
\begin{eqnarray}
&&\Big(\nabla^2\phi- \ft1{12}ae^{a\phi}G^2 -\ft18a e^{\ft12a \phi}
F^2\Big)\eta +\ft1{12}ae^{\fft12a\phi}\Gamma^{NM_1M_2M_3} (\nabla_N
G_{M_1M_2M_3}\cr && + \ft34 F_{NM_1} F_{M_2M_3})\eta
+\ft14ae^{-\fft12a\phi} \Gamma^{M_2M_3} \nabla_N\left(e^{a\phi}
G^N{}_{M_2M_3}\right)\eta=0\,,\label{intcon2}
\end{eqnarray}
and
\begin{equation}
\Gamma^{MM_1M_2} \hat D_{M} F_{M_1M_2} \eta + \Big[2
e^{-\fft12a\phi}\hat D^M(e^{\fft12a\phi} F_{MN})\eta + F^{M_1M_2}
G_{M_1M_2N}\Big]\Gamma^N\eta=0\,.\label{intcon3}
\end{equation}
Here, $\hat D\equiv d + [A,\,\,]$. In the above derivation, we have
used the important identity
\begin{equation}
\Gamma_{NM_1}{}^{M_2M_3} F^{NM_1} F_{M_2M_3} \eta =
\Big(\ft12\{\Gamma_{NM_1},\Gamma^{M_2M_3}\} + 2 \delta^{M_2}_{[N}
\delta^{M_3}_{M_1]}\Big)F^{NM_1}F_{M_2M_3}\eta = 2F^2\eta\,,
\end{equation}
which enables us to turn the $F_\2\wedge F_\2$ structure
from the Bianchi identity for $G_\3$ into the $F_\2^2$ term in the
Einstein equations of motion.  Analogously, we have used identity
$\Gamma_{M_1}{}^{M_2M_3} F_{M_2M_3}\eta = -2\Gamma^{M_2} F_{M_1M_2}
\eta$.  Thus the full set of equations of motion emerges from the
projected integrability conditions together with the Bianchi
identity for the 3-form. This demonstrates that the concept of the
Killing spinors is well defined in the effective theory, even with
the Yang-Mills fields. The results without the Yang-Mills were
previously obtained in \cite{lpw}.

As in the construction of extended supergravities, there is a
similarity between the curvature 2-form and the Yang-Mills fields,
such that the supersymmetrization of curvature-square term follows by
straightforward analogy with that for the Yang-Mills \cite{bss}. To
proceed, it is advantageous to work with the theory in the string frame,
which is defined by
\begin{equation}
ds^2_{\rm string} = e^{-\fft12 a\phi} ds_{\rm Einstein}^2\,.
\end{equation}
If we now define $\Phi=-\phi/a$, the Lagrangian becomes
\begin{equation}
{\cal L} = \sqrt{-g} e^{-2\Phi}(R + 4 (\partial\Phi)^2 - \ft{1}{12}
G_\3^2 -\ft14 {\rm tr}' F_\2^2)\,.
\end{equation}
The defining equations for the Killing spinors, which are scaled by
a factor $e^{-\fft18a\phi}$ in the string frame, are now given by
\begin{equation}
D_M(\omega_-) \eta=0\,,\qquad \Gamma^M \partial_M \Phi\,\eta
-\ft{1}{12} \Gamma^{MNP} G_{MNP} \eta=0\,,\qquad \Gamma^{M_1M_2}
F_{M_1M_2}\eta=0\,,\label{stringks}
\end{equation}
where $\omega_-$ is the torsional spin connection, defined as
\begin{equation}
\omega_{\mu\pm}{}^{ab} = \omega_{\mu}^{ab} \pm \ft12
G_{\mu}{}^{ab}\,.
\end{equation}
A major advantage of the string frame is that there is no manifest
dimensional dependence in either the action or the Killing spinor
equations. At the $\alpha'$ order, anomaly cancelation requires that
the quadratic curvature terms enter the Bianchi identity of the
3-form in the form of $dG_\3 \sim \ft12\alpha \Big({\rm
tr}(R_\2\wedge R_\2)-{\rm tr}' (F_\2\wedge F_\2)\Big)$, where
$R_\2=d\omega+ \omega\wedge \omega$ is the curvature 2-form and
${\rm tr} (R_\2\wedge R_\2)\equiv (R_\2)^{a}{}_b(R_\2)^b{}_a$. (Note
that for the ten-dimensional superstring, $\alpha=\ft12\alpha'$, and
we expect this to hold in general.) This suggests that we do not
need to modify the Killing spinors equations (\ref{stringks}), but
instead add an additional projection associated with the curvature
2-form {\it \`a la} the Yang-Mills. This projection in fact already
exists at the $\alpha'$ order, since \cite{bergroo}
\begin{equation}
0=[D_M(\omega_-), D_N(\omega_-)] \eta = \ft14
R_{MN}{}^{ab}(\omega_-) \Gamma_{ab}\eta = \ft14
R^{ab}{}_{MN}(\omega_+)\Gamma_{ab} \eta + {\cal
O}(\alpha)\,.\label{stringks1}
\end{equation}
This projection is analogous to that for the Yang-Mills fields
(\ref{ks3}), and hence the proper Bianchi identity for the 3-form is
given by
\begin{equation}
dG_\3 = \ft12\alpha  \Big({\rm tr}(R_\2(\omega_+)\wedge
R_\2(\omega_+))-{\rm tr}' (F_\2\wedge F_\2) \Big)\,.
\label{stringbianchi}
\end{equation}
Note that we adopt the supergravity convention that the torsionful
Riemann tensors are defined as $(R_\2(\omega_+))^a{}_b=\fft12
R_{MN}{}^a{}_b(\omega_+) dx^M \wedge dx^N$. Following the same
strategy as in the earlier calculation, we find that the full set of
equations of motion can now be obtained by the projected
integrability conditions, together with the Bianchi identity (\ref{stringbianchi}). They
are given by
\begin{eqnarray}
&&R-4(\partial\Phi)^2 + 4\Box\Phi - \ft1{12} G_\3^2 - \ft14\alpha
\Big( {\rm tr}' F_\2^2 - R_{MNAB}(\omega_+)
R^{MNAB}(\omega_+)\Big)=0\,,\cr 
&&R_{MN} + 2\nabla_M\nabla_N \Phi - \ft14 G_{MN}^2 - \ft12\alpha
\Big({\rm tr}' F_{MN}^2 - R_{MPAB}(\omega_+) R_{N}{}^{PAB}(\omega_+)
\Big)=0\,,\label{higherordereom}\\
&&d(e^{-2\Phi} {*G_\3})=0\,,\qquad \hat D(e^{-2\Phi}{*F_\2})+ (-1)^D
e^{-2\Phi} F_\2\wedge *{G_\3}=0\,. \nonumber
\end{eqnarray}
Up to the $\alpha'$ order, these equations can be derived from the
Lagrangian
\begin{equation}
{\cal L}_D = \sqrt{-g}e^{-2\Phi}\Big[R + 4(\partial\phi)^2 -
\ft1{12} G_\3^2 -\ft14\alpha \Big({\rm tr}'F_\2^2 -
R_{MNAB}(\omega_+)R^{MNAB}(\omega_+)\Big)\Big]\,.\label{alpha'lag}
\end{equation}
This Lagrangian in general dimensions takes the exact form of the
bosonic effective action of the heterotic string with
curvature-squard terms \cite{bergroo}.  The equations of motion
(\ref{higherordereom}) also take the same form as those in the
corresponding supergravity \cite{setbeck}. Note that the terms
arising from the variation of $\omega_+$ in the torsional Riemann
tensor do not appear in the equations of motion
(\ref{higherordereom}). These terms could at least contribute to the
$\alpha'$ order. As in the $R^2$ supergravity in $D=10$
\cite{bergroo}, at the $\alpha'$ order, we find that they
vanish by virtue of the leading-order equations of motion. Thus the
effective action of the bosonic string are now fixed, up to the
$\alpha'$ order, by the assumption that the Killing spinors can be
well defined. Conversely the above derivation also indicates
strongly that the pseudo-supersymmetry may be a property of the full
bosonic string.

    We now turn to the solutions of the bosonic string.  Let us
first set $\alpha'=0$. The theory admits the electric string
solution
\begin{eqnarray}
ds^2_{\rm str}&=& H^{-1}(-dt^2 + dx^2) + dr^2 + r^2
d\Omega_{D-3}^2\,,\cr
e^{2\Phi} &=& H^{-1}\,,\qquad F_\3 = dt\wedge dx\wedge
dH^{-1}\,,\qquad H=1 + \fft{Q}{r^{D-4}}\,,\label{stringsol}
\end{eqnarray}
and the magnetic $(D-5)$-brane
\begin{eqnarray}
ds^2_{\rm str}&=& \eta_{\mu\nu} dx^\mu dx^\nu + H (dr^2 + r^2
d\Omega_{3}^2)\,,\cr
e^{2\Phi} &=& H\,,\qquad e^{-2\Phi} {*F_\3}=dH^{-1} \wedge
d^{(D-4)}x\,, \qquad H=1 + \fft{P}{r^2}\,, \label{branesol}
\end{eqnarray}
and their intersection
\begin{eqnarray}
ds^2_{\rm str} &=& H_1^{-1} (-dt^2 + dx^2) + H_2 (dr^2 + r^2
d\Omega_3^2) + dx^i dx^i\,,\cr 
e^{2\Phi} &=& \fft{H_2}{H_1}\,,\qquad F_\3 = dt\wedge dx\wedge
dH^{-1} + e^{2\Phi} {*dt}\wedge dx\wedge d^{(D-6)} x\wedge
dH_2^{-1}\,,\cr 
H_1&=& 1 + \fft{Q}{r^2}\,,\qquad H_2 = 1 + \fft{P}{r^2}\,.
\label{stringbrane}
\end{eqnarray}
In the decoupling limit when the ``1'' in the harmonic functions
$H_i$ can be dropped, the metric (\ref{stringbrane}) becomes
AdS$_3\times S^3\times \mathbb{R}^{D-6}$.  As in supergravities, we shall
call these solutions ``BPS'' since they preserve fractions of the
maximally-allowed Killing spinors in the vacuum.

The Bianchi identity for $G_\3$ implies that the $(D-5)$-brane can
be supported by Yang-Mills instantons. For a single $SU(2)$
instanton, the Yang-Mills fields are given by $A_\1^i=a^2/(r^2 +
a^2)\,\sigma_i$, where $\sigma_i$ are the $SU(2)$ left-invariant
1-forms. The $S^3$ metric in (\ref{branesol}) can be expressed as
$d\Omega_3^2=\fft14 (\sigma_1^2 + \sigma_2^2 + \sigma_3^2)$.  The
$(D-5)$-brane takes the same form as (\ref{branesol}) but with $H$
now given by
\begin{equation}
H= 1 + \fft{r^2 + 2a^2}{(r^2 + a^2)^2}\,.
\end{equation}
This solution in the heterotic string was obtained in
\cite{stromingerhet}. For multi-instanton supported solutions, see
\cite{llop}.

     We now examine the $\alpha'$ corrected solutions.  It turns out
that, up to the $\alpha'$ order, (\ref{branesol}) is unmodified as a
solution for the Lagrangian (\ref{alpha'lag}).  This is because
the torsion for this solution is parallelizing and hence the curvature
2-form $R_\2(\omega_+)$ vanishes. If the existence of the Killing
spinors would restrict the higher-order corrections to be scalar
polynomials of $R_\2(\omega_+)$, the solution (\ref{branesol}) would
survive all the quantum corrections. This was first observed for
heterotic string in \cite{stromingerhet}, and we expect the same may
be true for the bosonic string with Killing spinors.

     In supergravities, the supersymmetry implies the existence of
the superspace geometry. The solutions can be classified by a
generalized holonomy. The generalized holonomy for the M-theory was
studied in \cite{duffliu,bdlw}. The existence of the well-defined
Killing spinors for the bosonic string suggests that the theory may
also have a generalized geometrical structure which can be called
pseudo-superspace.  The Killing spinor equations in (\ref{stringks})
imply that the generalized holonomy remains in the same $SO(1,D-1)$
as Einstein gravity since it merely adds a totally-antisymmetric
torsion to the usual spin connection. We now examine the reduced
holonomy for the $\ft12$-BPS solutions. We first look at the
electric string solution. We can replace the metric of the
$(D-2)$-dimensional transverse space by the one in the Cartesian
coordinates $dy^mdy^m$, in which $H$ is an arbitrary harmonic
function. For this background, the ``super''-covariant derivative is
given by
\begin{equation}
\mathcal{D}_{\mu}=\partial_{\mu} - \ft{1}{2} H^{-\fft32}\partial_m
H\,P^+\Gamma_{\hat \mu}{}^{\hat m} \,,\qquad \mathcal{D}_{m}=
\partial_{m} -\ft14H\partial_m H^{-1}\,\Gamma^{\hat0 \hat1} \,,
\end{equation}
where the projection operator is given by
$P^{\pm}=\ft12(1\pm\Gamma^{\hat0\hat1})$. Letting ${\mathcal
M}_{MN}=[\mathcal{D}_{M},\mathcal{D}_{N}]$, we find
\begin{equation}
{\mathcal M}_{\mu\nu}=0={\mathcal M}_{mn}\,,\qquad {\mathcal M}_{\mu
m}=\ft12H^{-\fft12}\partial_n\partial_m \ln H P^+\Gamma_{\hat
\mu}{}^{\hat n}\,.
\end{equation}
Since only the commuting generators $K_{\hat \mu}{}^{\hat
m}=P^+\Gamma_{\hat \mu}{}^{\hat m}$ are present, the reduced
holonomy for the electric string solution is given by
\begin{equation}
{\mathcal H}_{\rm string}=\mathbb{R}^{D-2}\,.
\end{equation}
The analysis for the magnetic $(D-5)$-brane is similar.  After
replacing the transverse space in (\ref{branesol}) with the
Cartesian system, the ``super''-covariant derivative is given by
\begin{equation}
\mathcal{D}_{\mu}=\partial_{\mu} \,,\qquad \mathcal{D}_{m}
=\partial_{m} +\ft12H^{-1}\partial_n H\,P^+\,\Gamma_{\hat m}{}^{\hat
n} \,,
\end{equation}
where the projection operator is given by $P^{\pm}= \ft12(1 \pm
\Gamma^{\hat1\hat2\hat3\hat4})$. We now have
\begin{eqnarray}
{\mathcal M}_{\mu\nu}&=&0\,,\qquad {\mathcal M}_{\mu m}=0\,,\cr 
{\mathcal M}_{mn} &=& \ft12\Big(\partial_{m}\partial_{n_1}
f-(\partial_{m} f)(\partial_{n_1} f)\Big)P^+\Gamma_{\hat n}{}^{\hat
n_1} -\ft12\Big(\partial_{n}\partial_{n_1} f-(\partial_{n}
f)(\partial_{n_1} f)\Big)P^+\Gamma_{\hat m}{}^{\hat n_1}\cr 
&& -\ft12(\partial^{n_1} f)(\partial_{n_1} f)P^+\Gamma_{\hat m\hat
n}\,,
\end{eqnarray}
where $f=\ln H$. The non-vanishing generators are $T_{\hat m\hat
n}=P^+\Gamma_{\hat m\hat n}$. Since
\begin{eqnarray}
\left[P^+\Gamma_{\hat m_1  m_2},P^+\Gamma_{\hat n_1 \hat
n_2}\right]=P^+\left[\Gamma_{\hat m_1  m_2},\Gamma_{\hat n_1 \hat
n_2}\right]\,,
\end{eqnarray}
it follows that $T_{\hat m\hat n}$ generates the $so(4)$ algebra.
Thus the reduced holonomy for the $(D-5)$-brane is given by
\begin{eqnarray}
{\mathcal H}_{\hbox{$(D-5)$-brane}}=SO(4)_+\,,
\end{eqnarray}
where $+$ refers to the sign of the $P_+$ projection.  For the
$\ft14$-BPS intersecting solution, we find that the reduced holonomy
is ${\mathcal H}=SO(4)_+\ltimes2\mathbb{R}^{(2_s)}$, where
$\mathbb{R}^{(2_s)}$ is the two-dimensional spinor representation of
the $SO(4)_+$.

   To conclude, we obtain the effective action for the bosonic string
up to the $\alpha'$ order. The guiding principle in the construction
is that the theory has well-defined Killing spinor equations, whose
projected integrability condition yields the full set of equations
of motion.  The success of our construction suggests that the hidden
pseudo-supersymmetry associated with the Killing spinor equations
may be a property of the full bosonic string. The
pseudo-supersymmetry enables us to classify solutions with respect
to the fractions of the surviving Killing spinors. These solutions
are characterized by the different reduced holonomy groups which are
subgroups of $SO(1,D-1)$. It is tempting to conjecture that these
hidden symmetry may cure some of the pathologies in the bosonic
string. Furthermore, as in the case of the superstring and M-theory,
the classical solutions with Killing spinors may also be protected
from quantum corrections and hence provide tools for studying the
non-perturbative aspects of the bosonic string.

\section*{Acknowledgement}

We are grateful to Haishan Liu, Yi Pang and Chris Pope for useful
discussions.

\end{document}